\documentclass[11pt,letterpaper]{article}

\usepackage[utf8]{inputenc}
\usepackage[T1]{fontenc}
\usepackage{amsmath,amssymb,amsfonts}
\usepackage{booktabs}
\usepackage{caption}
\usepackage{float}
\usepackage{graphicx}
\usepackage{xcolor}
\usepackage{enumitem}
\usepackage{setspace}
\usepackage[margin=1in]{geometry}
\usepackage{fancyhdr}
\usepackage{titlesec}
\usepackage{abstract}
\usepackage{hanging}   
\usepackage{hyperref}
\hypersetup{
  colorlinks=false,
  pdfborder={0 0 0},
  pdfauthor={Ajay Kumar Verma, Jul Jon Ramirez General, Yvan Landry Ndzonde Fonkou}
}

\titleformat{\section}{\large\bfseries}{\thesection.}{0.5em}{}
\titleformat{\subsection}{\normalsize\bfseries}{\thesubsection}{0.5em}{}
\titleformat{\paragraph}[runin]{\normalsize\bfseries}{}{0em}{}[.]

\usepackage{xcolor} 
\usepackage{datetime}
\title{From Classical Optimization to Bayesian Integration: A Comprehensive Analysis of Systematic Portfolio Management}
\newdateformat{monthyear}{\monthname[\THEMONTH] \THEYEAR}

\author{
\textcolor{gray}{
Ajay Kumar Verma$^{1}$, 
Shravya Barkam$^{2}$
}
}

\begin{document}

\date{\monthyear\today}

\maketitle

\renewcommand{\thefootnote}{}

\footnotetext{
\small
$^{1}$Independent Researcher (azay.verma@gmail.com)
 | 
$^{2}$Independent Researcher (shravya.barkam@gmail.com)
}

\renewcommand{\thefootnote}{\arabic{footnote}}

\begin{abstract}

This paper compares a series of contemporary portfolio construction approaches by employing ten U.S. stocks (TSLA, WMT, BAC, GS, LLY, MRK, GOOG, META, AAPL and XOM) in a time frame from September 2023 to December 2025. The paper explores both basic mean-variance optimization, constrained optimization, Fama French five factor regression modeling, Monte Carlo simulation, and the Black-Litterman model to determine how constraints to a solution, risk factors to a strategy, simulated approximations, and specific market views may all impact the outcome of portfolio allocation, performance and stability. Overall, the results show that standard optimization may result in highly concentrated portfolios, while constrained optimization leads to changes in portfolio allocations by altering the efficient frontier, five factor regression models suggest that a basic investment style of defensive large value and profitability exposure, Monte Carlo approximation is a viable technique to arrive at mean-variance optimal portfolios provided the simulations are high enough especially under a box constraint, the Black Litterman portfolio approach produces more economically intuitive allocations and greater stability compared to standard mean-variance optimization as the approach balances equilibrium returns with investor views.

\end{abstract}

\newpage

\tableofcontents

\newpage

\section{Introduction}

The problem of portfolio optimization has been a staple in the quantitative finance research community since the development of the Modern Portfolio Theory by Markowitz in 1952. Since then, investors have aimed to generate portfolios with maximum expected return and lowest possible risk given their level of risk tolerance. This mean-variance framework allows practitioners to construct portfolios that minimize the risk at a given level of expected return based on the covariance between different securities as well as the risk of individual assets (Markowitz). 

However, mean-variance optimization in practice is often difficult to apply directly to an optimization problem, since it can be highly sensitive to expected returns and result in poorly diversified portfolios which might not make economic sense. Thus, it is common for investors to limit the possible efficient portfolios by setting some additional constraints such as restricting a portfolio to be long only, or adding upper bounds on the position size. Adding these constraints will reduce the attainable level of diversification for that portfolio. Furthermore, researchers have found a set of systematic risk factors that explain much of the portfolio return variation. For example, the Fama-French Five-Factor model extends the single factor CAPM by additionally incorporating size, value, profitability, and investment factors (Fama and French). 

These factors not only give investors an indication of the style and characteristics of their portfolio, they also serve as a way of explaining the cross-sectional difference in portfolio returns over time. Monte Carlo simulation has also been used to solve various types of optimization problems. A simulation of a Monte Carlo method can allow one to obtain a probabilistic approximation of the optimal portfolios when the dimensionality or number of constraints makes it impossible to solve using traditional analytic optimization techniques. The convergence of random search is also a function of the dimensionality of the parameter being optimized and the number of computations. The Black-Litterman model has also been introduced in this paper as another possible method of portfolio optimization that does not rely as heavily on the stability of expected returns to form an optimal portfolio. It combines equilibrium returns with investor views to form optimal portfolios which are less likely to be concentrated in specific stocks (Black Litterman).

This paper brings together all these methods in a single empirical framework. Through out-of-sample tests on a population of ten U.S. stocks, it analyzes how constraints, factor loadings, simulations and prior beliefs impact portfolio composition and choices made by investors.
\newpage
\section{Literature Review}

Markowitz’s contribution to mean-variance optimization theory, which treats portfolio diversification as an optimization problem (Markowitz), forms the theoretical basis of portfolio optimization. The set of best-performing portfolios for any amount of risk is known as the efficient frontier. Sharpe built on this theory with his work to develop the capital asset pricing model (CAPM), which links the expected returns of a portfolio to market risk (represented as a beta exposure) (Sharpe). While mean-variance optimization has always been at the center of institutional portfolio management, research has shown its limitations: the resulting portfolios are often highly unstable due to measurement errors and are sensitive to small changes in return forecasts, which can cause substantial changes in portfolio weights (Michaud 31). The optimal weight vectors generated by the mean-variance optimization technique can vary considerably with minor perturbations in return forecasts, which leads to corner portfolios that have highly concentrated exposures to the asset class. Later research by Fama and French sought to better capture the systematic sources of returns and risk by augmenting the model with other market factors. In their work, Fama and French present a five-factor model that incorporates, along with the equity premium and market beta, size (SMB), value (HML), profitability (RMW), and investment (CMA) factors (Fama and French). They find that exposure to SMB and HML explains the returns of stocks across many market environments and over the long time horizons. While the mean-variance optimization methodology has been shown to be highly sensitive to the expected returns of each asset, research by Fama and French suggests that the size, value, and style of portfolio managers have significant influence on portfolio performance and that the exposure of these risk factors varies over time. Monte Carlo optimization, which involves creating a large number of random admissible portfolio realizations and selecting the optimal based on a risk-return trade-off, is increasingly employed to solve portfolio optimization problems (Glasserman). Monte Carlo simulation methods are flexible and intuitive, which means they can be adapted to solve difficult optimization problems. However, Monte Carlo optimization is not without its limitations, and the curse of dimensionality can significantly reduce the computational efficiency in constrained problem instances. Developed at Goldman Sachs, the Black-Litterman model is based on Bayes’ theorem. It combines prior expectations of asset returns with views about future performance, producing a new set of posterior expectations (Black and Litterman). This model is a significant improvement over classical mean-variance portfolio optimization approaches, which rely on the expected rate of returns based on historical data. By combining the equilibrium mean returns with the investor’s views, the Black-Litterman method generates more diversified portfolios and stabilizes the estimates of asset returns for the portfolio manager.

Recent research reinforces the importance of optimization robustness, behavioral portfolio construction, and factor diversification in asset management. Portfolio managers often pair optimization with economic and macro views, and increasingly with methods like machine learning, to enhance robustness in the face of uncertainty (Meucci).

\section{Data and Methodology}

\subsection{Data Description}

The study examines the following ten securities:

\begin{center}
TSLA, WMT, BAC, GS, LLY, MRK, GOOG, META, AAPL, XOM
\end{center}

To compute the returns, we retrieved the adjusted daily closing stock prices of the underlying assets on Yahoo Finance for the estimation period of September 1, 2023 to September 30, 2025. Furthermore, the out-of-sample evaluation period of October 1, 2025 to December 31, 2025, was adopted to assess the performance of all constructed portfolios. 

Returns were computed using logarithmic differences.

\[
r_t = \ln \left(\frac{P_t}{P_{t-1}}\right)
\]

where \(P_t\) denotes the adjusted closing price at time \(t\).

Annualized expected returns and covariance matrices were estimated using:

\[
\mu = 252 \times \bar{r}
\]

\[
\Sigma = 252 \times Cov(r)
\]

where \(252\) represents the approximate number of trading days in a calendar year.

\section{Mean-Variance Portfolio Optimization}

\subsection{Unconstrained MVO and the Corner Portfolio Problem}

The initial optimization framework implemented in this study solves for the \textit{Global Minimum Variance (GMV)} a portfolio which is not necessarily the tangent portfolio. Whereas the tangent portfolio gives the Sharpe maximising portfolio as it simultaneously minimizes the volatility and maximizes the expected return, the optimization problem solved in this work does not take the expected return into consideration but rather focuses on the variance minimization as long as it satisfies certain requirements.

Formally, the optimization problem is specified as:

\[
\min_{w} \quad w^{T}\Sigma w
\]

subject to:

\[
\sum_{i=1}^{n} w_i = 1
\]

\[
w_i \geq 0
\]

\[
w_i \leq w_{\max}
\]

where \(w\) denotes the vector of portfolio weights, \(\Sigma\) represents the covariance matrix of asset returns, and \(w_{\max}\) aligns to the maximum amount that can be allocated to any individual asset; in this case, this is 1

This objective minimizes the total variance of the portfolio and does not include a direct return term. Therefore, the solution corresponds to the minimum variance portfolio within the defined long only and box constraints. It does not correspond to an efficient frontier or tangency portfolio, which take into account both the return and the risk of the portfolio. 

This optimization was performed using quadratic programming through the \texttt{cvxpy} library. Quadratic programming is particularly appropriate in this setting because the variance term \(w^{T}\Sigma w\) is quadratic in portfolio weights while the investment constraints remain linear.

The unconstrained optimization, which only required weights to sum to one and prohibited short-selling, identified a highly concentrated "corner portfolio" dominated by AAPL (29.05\%), XOM (27.83\%), and GOOG (21.75\%). The unbounded global minimum variance (GMV) portfolio is heavily concentrated in relatively stable, large-cap assets with positive covariance profiles. That is, those assets delivered the strongest relative diversification benefits per unit of idiosyncratic and pairwise risks.

A notable feature of unconstrained minimum variance optimization is that the resulting asset weights are solely a function of the covariance structure and not the expected returns; as such, securities with low marginal variance contribution dominate the allocation even if they are not the highest expected return securities (this is the reason for the large AAPL and XOM weights despite the high expected return securities such as TSLA and META). The optimal portfolio obtained from this optimization problem reflects the bias of minimizing variance portfolios towards lower overall portfolio variance. As we will explore later, this bias can be overcome by imposing constraints on individual portfolio weights, or by optimizing a variance-adjusted return proxy; however, as is evident from our current results, unconstrained minimum variance portfolios can sacrifice considerable returns for the reduced aggregate portfolio variance.

\begin{table}[H]
\centering
\caption{Unconstrained Global Minimum Variance Portfolio Weights}
\label{tab:gmv_weights}
\begin{tabular}{l c}
\toprule
\textbf{Asset} & \textbf{Optimal Weight} \\
\midrule
TSLA & 0.0076 \\
WMT  & 0.0389 \\
BAC  & 0.1335 \\
GS   & 0.0000 \\
LLY  & 0.0301 \\
MRK  & 0.0036 \\
GOOG & 0.2175 \\
META & 0.0000 \\
AAPL & 0.2905 \\
XOM  & 0.2783 \\
\midrule
\textbf{Total} & \textbf{1.0000} \\
\bottomrule
\end{tabular}
\end{table}

\subsection{Constrained Optimization : The Impact of Diversification Constraints}

To fulfill the institutions requirements, a further box restriction was included, requiring the portfolio to be more diversified, with each individual asset being no more than 15 percentage:

\[
w_i \leq 0.15
\]

\begin{table}[H]
\centering
\caption{Constrained Global Minimum Variance Portfolio Weights (15\% Maximum Allocation)}
\label{tab:constrained_gmv_weights}
\begin{tabular}{l c}
\toprule
\textbf{Asset} & \textbf{Optimal Weight} \\
\midrule
TSLA & 0.1215 \\
WMT  & 0.1500 \\
BAC  & 0.1500 \\
GS   & 0.0058 \\
LLY  & 0.1057 \\
MRK  & 0.0170 \\
GOOG & 0.1500 \\
META & 0.0000 \\
AAPL & 0.1500 \\
XOM  & 0.1500 \\
\midrule
\textbf{Total} & \textbf{1.0000} \\
\bottomrule
\end{tabular}
\end{table}

This constrained Global Minimum Variance (GMV) portfolio illustrates the importance of diversification when constructing a portfolio. Since this constraint prevented the maximization from allocating such a high percentage of weights to just a small number of securities in a portfolio, there was significant diversification of allocation among a range of securities. WMT, BAC, GOOG, AAPL, and XOM were at or very near the constraint (15\%), suggesting that the optimizer would have liked to place even more weight in them if allowed. In effect, these securities are the most attractive variance diversifiers available in the portfolio once accounting for the covariation in the portfolio. TSLA and LLY obtained fairly decent allocations, but did not quite meet the constraint (at 12.15\% and 10.57\%). These assets still provide diversification benefits but due to their higher volatility, could not attain the maximum bound allowed to other assets. MRK and GS only received relatively low allocations while META only received a 0\% allocation. Therefore these securities provided little to no diversification benefits. Unlike the unconstrained portfolio, we have a relatively equal weight in a much larger number of stocks while still significantly reducing variance risk. Yet, because this constraint prevents the optimizer from assigning higher weights to securities that can lower the variance, we end up being pushed away from our efficient frontier. As can be observed economically, in theory an unconstrained optimization portfolio can minimize portfolio variance more effectively than a constrained portfolio. However, institutions, such as fund managers, often restrict the allocation percentages to individual stocks and thus reduce idiosyncratic risk and increase diversification.

\begin{figure}[h!]
    \centering
    \includegraphics[width=1\linewidth]{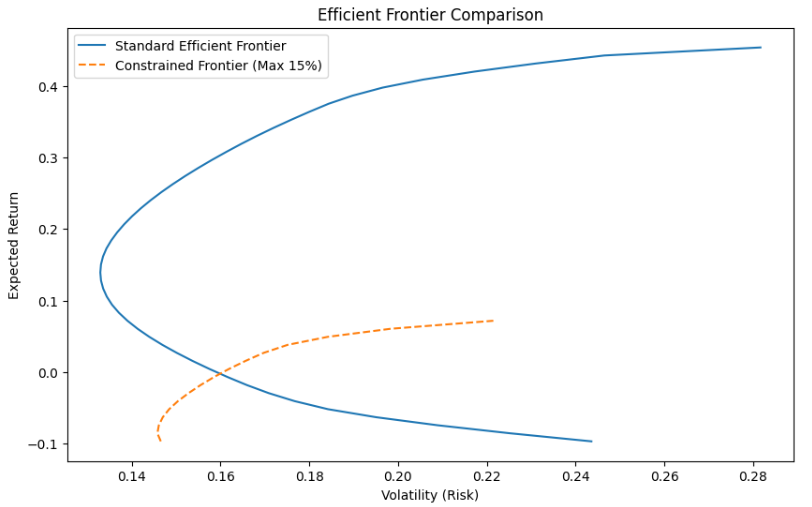}
    \caption{Constrained vs Unconstrained Efficient Frontier}
    \label{fig:placeholder}
\end{figure}

Here are a few key conclusions we can draw from comparing the unconstrained and constrained frontiers and understanding how constraints affect the feasible opportunity set for a diversified portfolio.

Firstly, the constrained frontier is relatively less efficient than the unconstrained frontier. For any given volatility, constrained portfolios exhibit less expected returns. This is because the weights are restricted, so the optimization algorithm cannot assign larger weights to the historical out-performers. Hence it can't find the optimal asset combinations. Secondly, the constrained frontier covers less space on the mean-variance space. This makes the range of feasible expected returns and volatility smaller. In contrast, the unconstrained frontier permits highly concentrated allocations, including extreme corner solutions in which the optimizer may allocate nearly the entire portfolio to a single high-performing asset. In that sense, the unconstrained frontier has better possible maximum returns, however it is often achieved by much higher risks. 

Thirdly, the diversification rule causes a sub-optimal expected risk-return relationship from purely mathematical perspective. On one hand, the weights cap increases diversification and reduces concentration risk. On the other hand, it requires holding asset combinations which are sub-optimal on the mean-variance efficiency metric. Thus the manager trades some of the efficiency for more diversification and better implementability. Overall, with box constraints on the weights, the feasible space for optimization becomes smaller and the efficient frontier is now lower and to the right. This proves that constrained portfolios are strictly less efficient than unconstrained portfolios on the classical Markowitz efficient frontier model.

\newpage
\subsection{Out-of-Sample Performance Evaluation}

In the out-of-sample period of October to December 2025, performance of the portfolio was assessed on the basis of cumulative return, the Sharpe ratio and maximum drawdown.

The Sharpe ratio was computed as:

\[
SR = \frac{R_p - R_f}{\sigma_p}
\]

where \(R_p\) denotes portfolio return, \(R_f\) represents the risk-free rate, and \(\sigma_p\) is portfolio volatility. This measure gauges whether the portfolio manager's active approach created value on a risk-adjusted basis for the time period under scrutiny. 

Maximum drawdown, often described as “the maximum possible loss,” measures a client’s actual exposure to risk by calculating the largest decline from a high point to a low point within a specific timeframe.

\begin{table}[H]
\centering
\caption{Out-of-Sample Portfolio Performance (October 2025 -- December 2025)}
\label{tab:oos_performance}
\begin{tabular}{lccc}
\toprule
\textbf{Portfolio} & \textbf{Cumulative Return} & \textbf{Sharpe Ratio} & \textbf{Maximum Drawdown} \\
\midrule
Unconstrained GMV Portfolio & 15.85\% & 5.05 & -2.76\% \\
Constrained GMV Portfolio   & 16.77\% & 5.36 & -3.18\% \\
\bottomrule
\end{tabular}
\end{table}

We turn next to out-of-sample results, which are an important measure of the portfolio's robustness during the sample period from October 2025 to December 2025. Surprisingly, the constrained portfolio outperformed the unconstrained portfolio with regard to both the cumulative return and the Sharpe ratio, in spite of the loss in optimization efficiency associated with adding the diversification constraints. The constrained portfolio earned a cumulative return of 16.77\%, slightly more than the 15.85\% earned by the unconstrained portfolio. Moreover, the Sharpe ratio for the constrained portfolio was 5.36, compared to 5.05 for the unconstrained portfolio. These results indicate that the diversification constraints increased the portfolio stability and robustness during the out-of-sample evaluation. The constrained portfolio is likely more stable and robust because the diversification constraints prevent the portfolio from being overly concentrated in only a few securities, thereby reducing sensitivity to estimation error and idiosyncratic risks. In addition to the cumulative return and the Sharpe ratio, we report the maximum drawdown. 

The constrained portfolio has a slightly larger maximum drawdown of -3.18\%, whereas the unconstrained portfolio has a maximum drawdown of -2.76\%. The constrained portfolio is therefore slightly more sensitive to short-term losses, though the difference in the maximum drawdown between the two is rather small. In this case, the maximum drawdown of the constrained portfolio indicates that, on a short-term basis, it experienced marginally larger losses than did the unconstrained portfolio. The out-of-sample results offer a practical illustration of one of the limitations of optimization. Although the unconstrained portfolio has a higher return and higher Sharpe ratio in-sample, it performs poorly during the out-of-sample evaluation. In general, the unconstrained portfolio, while more efficient in-sample, may be less robust out-of-sample because of the high sensitivity to estimation error and the potential for a regime shift in the market. In contrast, the constrained portfolio appears to be slightly more stable and robust out-of-sample, as the portfolio is diversified enough to prevent overly concentrated allocations and to ensure better portfolio performance under various market conditions. 

This case study suggests that in the presence of reasonable diversification constraints, the actual out-of-sample portfolio return may be better, even if the constrained portfolio theoretically does not lie on the efficient frontier in the context of mean-variance analysis.

\section{Fama-French Five-Factor Analysis}

We then took the resulting portfolio from the constrained optimization part and ran it through the Fama-French Five-Factor Model:

\[
E[R_p]-R_f = \alpha + \beta_1(R_m-R_f)+\beta_2SMB+\beta_3HML+\beta_4RMW+\beta_5CMA+\epsilon
\]

The five factors represent:

\begin{itemize} \item Market Risk Premium (\(R_m-R_f\)) : This factor reflects the portfolio’s "systematic" risk exposure, or its sensitivity to movements in the overall stock market. \item SMB (Small Minus Big) : Called the Size Factor, SMB is the historical risk premium on small companies relative to large companies; a higher coefficient means the portfolio has a bias toward small-cap stocks. \item HML (High Minus Low) : Known as the Value Factor, HML refers to the historical premium on “high” (value) stocks, which have high book-to-market ratios, over low (growth) stocks, which have lower book-to-market ratios. \item RMW (Robust Minus Weak) : This captures the profitability factor, which reflects the difference between the average returns of stocks with robust (high) operating profitability versus stocks with weak (low) operating profitability. \item CMA (Conservative Minus Aggressive) : This denotes the investment factor, which is the difference between the average returns of companies that invest more conservatively versus companies that invest more aggressively. \end{itemize}

\subsection{Factor Correlations}

\begin{table}[H]
\centering
\caption{Fama-French Five-Factor Correlation Matrix}
\label{tab:ff5_corr}
\begin{tabular}{lccccc}
\toprule
 & \textbf{Mkt-RF} & \textbf{SMB} & \textbf{HML} & \textbf{RMW} & \textbf{CMA} \\
\midrule
\textbf{Mkt-RF} & 1.000 & 0.219 & -0.269 & -0.410 & -0.027 \\
\textbf{SMB}    & 0.219 & 1.000 & 0.378 & -0.395 & 0.174 \\
\textbf{HML}    & -0.269 & 0.378 & 1.000 & 0.214 & 0.189 \\
\textbf{RMW}    & -0.410 & -0.395 & 0.214 & 1.000 & -0.012 \\
\textbf{CMA}    & -0.027 & 0.174 & 0.189 & -0.012 & 1.000 \\
\bottomrule
\end{tabular}
\end{table}

The Fama-French five factors’ correlation matrix shows that they generally have low to moderate correlation, which is one of the key reasons we use multi-factor models for equity asset pricing. The market risk premium factor (Mkt-RF) has a low positive correlation with the size factor (SMB) (0.219). This suggests that smaller firms slightly outperformed larger firms during the sample period when the market performed well. However, the market factor negatively correlated with the value factor (HML) and profitability factor (RMW), which has a stronger negative correlation (-0.410). This finding suggests that high-profit firms may exhibit defensive characteristics, which could be helpful in diversifying portfolios in weak market conditions. The size factor has moderate positive correlation with the value factor (0.378). This indicates that small firms in the sample period had value-like qualities, such as higher book-to-market ratio. On the other hand, small firms appear to be negatively correlated with profitability. The profitability (RMW) factor is nearly uncorrelated with the investment factor (CMA) with a correlation coefficient close to zero (-0.012). This weak correlation implies that the profitability factor and the CMA capture different types of behavior of firms. Importantly, no off-diagonal coefficients are so large that we would be concerned about multicollinearity. Hence, the regression coefficients should be interpretable. In general, low inter-factor correlations suggest benefits from factor diversification.

Overall, the correlation structure confirms the theoretical justification of the Fama-French approach that the five factors represent five unique sources of systematic risk rather than simply different measures of the same single source of risk.

\subsection{Regression Analysis}

OLS, the most common method for estimation of factor loadings. For the model, OLS estimation requires the assumption that the errors in the model are homoskedastic (i.e., all error terms have the same variance) and that residuals are normally distributed. Robust regression methods, on the other hand, are included in this study to account for the fact that the distributions of financial returns have very heavy tails and very large outliers that are inconsistent with the classical assumptions made in the OLS approach. By down-weighting such outlier observations, the robust model is likely to produce a better estimate of the "typical" level of exposure to the factors.

\subsection{Data Division and Splitting Strategy}

We split the sample in two subperiods to be used, respectively, to estimate the model and to test it, which allows a more rigorous out-of-sample test without look-ahead bias. 

\subsubsection{Training Data (Estimation Period)}

The estimation period runs from September 1, 2023 through September 30, 2025. Using the data in this period, the expected returns, variance-covariance matrix, and factor sensitivities were estimated, that is, the period when the regression models of Fama and French’s five-factor model ``learn'' what the factor exposures of the portfolio have been historically. More specifically, the Ordinary Least Squares (OLS) as well as Robust Linear Model (RLM) regressions were run over this time to estimate the sensitivities of the portfolio to the following factors:

\[
(R_m - R_f), \quad SMB, \quad HML, \quad RMW, \quad CMA
\]

The estimated betas represent the portfolio's historical style and systematic risk. These coefficients describe the riskiness of the optimized portfolio. By estimating models on the training sample only, we avoid contamination of out-of-sample test.

\subsubsection{Testing Data (Out-of-Sample Period)}

The second period of interest extends from October 1st, 2025 to December 31st, 2025 and is used for testing and for the evaluation of out-of-sample results. Results for this time period were not included in the optimization or regression estimation processes so that the results are not biased. The objective of the testing period is to observe whether the portfolio characteristics and factors remain consistent and if the results are economically feasible. It allows for a more realistic assessment of portfolio consistency and reliability. Results for the testing period are assessed based on standard portfolio performance metrics, namely, cumulative return, the Sharpe Ratio, and Maximum Drawdown.

\subsubsection{Rationale for the Splitting Strategy}

This split between the estimation and test samples is necessary to avoid look-ahead bias and overfitting. If we used the same data to optimize the portfolios and compute the statistics, our results would likely reflect the over-optimization of the process rather than the true in- and out-of-sample predictive power. In addition, keeping the sample and the test sample in the same frequency helps us obtain comparable results when estimating covariances and factor exposures. Using a pure sample period for testing (last quarter in 2025) allows us to see how the optimized portfolios actually perform rather than how well we fit the sample data and also provides a more realistic scenario that is often used by institutions when back-testing portfolio strategies.

\subsection{Interpretation of OLS Regression Results}
\begin{figure}[h!]
    \centering
    \includegraphics[width=1\linewidth]{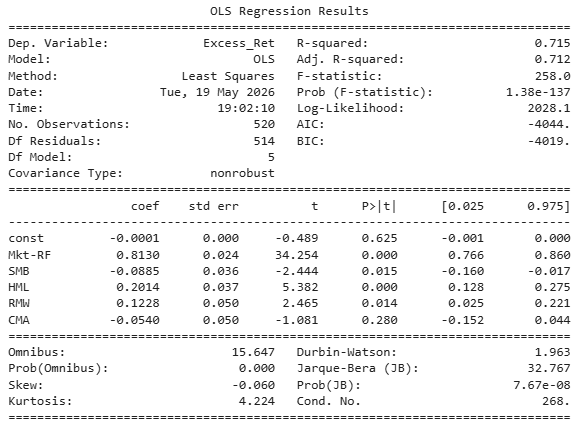}
    \caption{OLS Regression Results}
    \label{fig:placeholder}
\end{figure}

Table shows OLS regression estimates of Fama and French (2015) five-factor model, where dependent variable is excess returns from optimized portfolio. The result from the table indicate a very high model fit. In fact the \(R^2\) value is 0.715 and the adjusted \(R^2\) value is 0.712. That is the explanatory power is high. This means that 71.5\% of the movements of the excess returns of the optimized portfolio is explained by the five factors. The \(F\) statistic is 258.0 which indicates that there is a significance between the variables. Moreover, p-value is smaller than 0.001 which confirms our findings. In other words, the five factors explain the optimized portfolio’s risk premium. The intercept term is \(\alpha\) is estimated by -0.0001. The p-value is 0.625 which is statistically insignificant. In other words, the value of the alpha is small which indicates that the optimized portfolio does not achieve a greater return compared to the systematic factors. This finding implies that this portfolio does not outperform the factor model. The absence of statistically significant alpha is broadly consistent with efficient market theory. Additionally, the value of \(Mkt-RF\) factor is significant since the p-value smaller than 0.001. The result shows that the beta coefficient of \(Mkt-RF\) is 0.8130 which indicates that the excess return of the optimized portfolio is affected by the market. This positive coefficient means that excess return of this portfolio will also increase (decrease) when the market return increases (decreases). However, since the \(Mkt-RF\) beta coefficient is smaller than 1 this portfolio is less sensitive to risk factor compared to the market which indicates that it has a low amount of exposure to risk factor.

Size is found to be negative and statistically significant at \(-0.0885\) ( \(p = 0.015\) ), which implies a tilt towards large cap stocks instead of small cap stocks. The large cap firms tend to be financially more stable and have higher liquidities and hence less operating risks, which align with the defensive nature captured by the negative market beta. Value is positive and statistically significant at \(0.2014\) ( \(p < 0.001\) ), which suggests that there exists a large exposure to value stocks (i.e. high book-to-market ratio firms) in the portfolio. Value stocks tend to be mature, profitable companies with more stable earnings and relatively high cash flows, and have lower valuation multiples. Profitability is also positive and statistically significant at \(0.1228\) ( \(p = 0.014\) ), which indicates an inclination for firms with high operating profitability in the portfolio. Investment is negative but statistically insignificant ( \(p = 0.280\) ), which means we are unable to conclude the portfolio has significant exposures to firms with conservative or aggressive investment policies. Durbin-Watson of \(1.963\) is close to \(2\) , which suggest no substantial amount of serial correlations in the error terms. This implies that the model errors are roughly time independent. The Jarque-Bera and Kurtosis measures imply slight departure from the normality assumption of the errors, where the kurtosis (measured by fourth central moment) of 4.224 > 3 , suggesting mildly fat tails in the residual distribution. This is not uncommon in financial time series, and motivates subsequent robust regression.

Overall, the OLS regression results characterize the optimized portfolio as a defensive large-cap value portfolio with meaningful exposure to very profitable firms. The factor structure captures a significant portion of portfolio return volatility, showing the importance of systematic risk factors in explaining the characteristics and style of this particular portfolio.

\subsection{Interpretation of Robust Regression Results}

The table below reports the regression coefficients from a Robust Linear Model (RLM) using the HuberT norm and Iteratively Reweighted Least Squares (IRLS). Given that financial data are often characterized by fat tails, volatility, and potential outliers which can heavily distort the OLS estimates, robust regression is the preferable approach. This method mitigates the undue influence of outliers and provides a more stable estimate of the portfolio's average exposures. Consistent with the OLS regression above, the estimates from the robust regression are close to their OLS counterparts, suggesting that the overall factor structure is not being driven by extreme observations in the sample. This strengthens confidence in the estimates. As in the OLS regression, the alpha coefficient is insignificant (p = 0.952). The absence of a significant alpha implies that the portfolio does not outperform the factor model on a risk-adjusted basis. This result is consistent with efficient market theory and suggests that returns on the portfolio are well explained by market-wide factors. Again, Mkt-RF is the most significant factor, with a positive coefficient of 0.7970 significant at the 1

We still find a statistically significantly negative (SMB) loading, equal to -0.0728 ( p = 0.036 ). This means the portfolio has a large cap tilt even when we remove the effects of outliers. In comparison to the OLS results, the magnitude of (SMB) is slightly less, which implies some outliers may have contributed to the strength of the large-cap tilt under OLS. The coefficient of (HML) is 0.1827, with a p-value of less than 0.001. The positive and highly significant coefficient of (HML) indicates that there is an exposure to value firms in the portfolio, meaning value is a component of returns in the portfolio. Additionally, the coefficient of (RMW) in the robust regression is 0.1507, with a p-value of 0.002. We see that the coefficient of (RMW) increases from that under the OLS regression, implying the return on profitable firms contributes to the return of the portfolio and the effect is magnified when outliers are discounted. The p-value of (CMA) is 0.473. The coefficient of (CMA) is insignificant, meaning it cannot be said with certainty that the portfolio is skewed towards conservative or aggressive firms. A significant aspect of using the robust regression is the model’s ability to account for heavy tails of the distribution that are frequently found in financial data. The HuberT norm is applied in order to reduce the effect of outliers on the estimation of regression parameters; it is accomplished by assigning low weights to outliers in the iteration of the least squares. Thus, the robust regression is applicable to the financial returns that are often heavy-tailed, and do not necessarily have to follow a normal distribution.

In short, the robust regression results broadly reaffirm the OLS findings. This portfolio is a defensive large-cap value portfolio that has a high loading to high profitability. The close similarity between the OLS and RLM estimates indicates that the factor exposures are not too influenced by extreme observations or temporary shocks.

\begin{figure}[h!]
    \centering
    \includegraphics[width=1\linewidth]{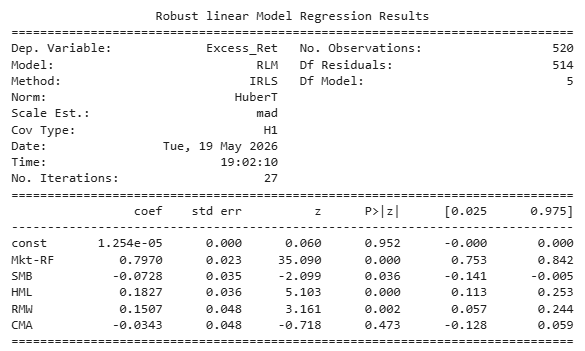}
    \caption{Robust Regression Results}
    \label{fig:placeholder}
\end{figure}

\section{Factor Exposure Analysis of the Constrained Portfolio}

As the regression results reveal, a significant fraction of the constrained portfolio’s excess return variation is explained by the Fama-French factors. The \(R^{2}\) of the OLS regression is greater than 70\%. In economic terms, this indicates that variation in the constrained portfolio’s returns results from risk exposure to systematic risk factors, as opposed to the effects of a specific stock on its return. The OLS and robust regression \(\alpha\) is not significantly different from 0; economically, we interpret this result to mean that the constrained portfolio does not earn excess returns once risk exposures to the Fama-French factors are taken into account. Thus, variation in the constrained portfolio’s return reflects returns to exposure to market and style-based risk factors.

\subsection{Market Risk Exposure}

The regression results identify the market factor (\textit{Mkt-RF}) as the predominant explanatory variable. Furthermore, the market beta estimate is less than 1 for the OLS and robust regression, suggesting that the constrained portfolio has defensive nature. Specifically, the portfolio beta implies that the portfolio is aligned with the market movement, yet more conservative than the market in general (i.e., if the market moves up by 1\%, the portfolio moves up by 0.75 to 0.8\% on average). This implies that the portfolio is less volatile than the market in general. We expect that this result is driven by the diversification constraints imposed on the portfolio optimization procedure, such that the optimization will only include securities with low contribution to the portfolio's total risk.

\subsection{Size Factor Exposure}

The Size factor (\textit{SMB}) is loaded negatively in both regressions, indicating that the constrained portfolio exhibits a large-cap bias rather than exposure to small-cap firms. Economically, this means that the portfolio has a negative exposure to SMB, implying a tilt towards more established companies such as AAPL, XOM, WMT, and GOOG. Large-cap companies tend to have stronger balance sheets, better liquidity, and less business uncertainty compared with smaller companies; thus, they offer more downside protection during periods of market turbulence. This large-cap tilt is consistent with the defensive nature of the portfolio, which, in return, yields a relatively stable out-of-sample performance observed during the testing period.

\subsection{Value Factor Exposure}

The value factor (\textit{HML}) is also statistically significant, with positive coefficient, for both regression methods. This implies that the restricted portfolio has a significant exposure to value stocks. Positive exposure to the HML factor implies that the portfolio favors stocks with relatively higher book to market ratios, smaller multiples of valuations, and older firms. Value stocks are often characterized by having consistent returns, stable market share and low speculative pricing. The significance of the value factor in the regression results suggests that the portfolio’s return is largely explained by factor exposure to value stocks, rather than growth oriented stock.

\subsection{Profitability Factor Exposure}

The profitability factor (RMW) also entered positively and significantly in both the OLS and robust regressions. This implies that the constrained portfolio is indeed tilted toward firms with higher operating profitability. Because higher profits typically imply superior quality of earnings, superior operational performance and improved balance sheet strength, it should come as no surprise that within this constrained portfolio of companies the companies with higher profits have historically generated excess returns in volatile periods, when times are more uncertain. These higher-profitable companies are less susceptible to downturns, and have greater free cash available to deploy in the future, which may be particularly useful when times are rough. Intriguingly, the profitability coefficient is even higher in the robust regression model, compared to the OLS model. This implies that the excess returns we earn from tilting our portfolios toward higher-profit companies survive our down-weighting of extreme, and thus presumably more volatile, return data points. Overall, the exposure to profitability within a portfolio adds further diversification in volatile environments, which can be very important for mitigating risk.

\subsection{Investment Factor Exposure}

The investment factor (CMA) is not significant under either specification. In other words, there is little evidence that our constrained portfolio tilts toward firms that are conservatively invested or firms that aggressively expand. Thus, it is not clear that corporate investment plays a major role in the return behavior of the portfolio over the estimation period. Market, value, and profitability exposures remain the dominant features of our constrained portfolio.

\subsection{Role of Factor Diversification}

One of the notable results from the factor analysis is the relatively small correlation among the factors. Specifically, the correlation between market and profitability is slightly negative, so profitable firms tend to be more defensive during downturns. This diversification benefit is especially important because, unlike asset classes, style factors often remain uncorrelated during stress. This helps the systematic risk factors provide a relatively stable risk premium across various regimes and conditions. Therefore, exposure to defensive market factors, big companies, value, and profitability provides good diversification for the systematic factors.

\subsection{Practical Investment Implications}

Practically, the constrained portfolio appears to serve the needs of an investor who has a preference for lower levels of risk and would like a reliable source of long-term returns, while protecting against excessive risk of loss. The portfolio’s beta, which is below 1, and its exposure to the value factor and to high profitability firms, match the investor desire to invest in high-quality, well-established firms, which will generate adequate free cash flow under adverse economic conditions. These results demonstrate the need to use alternative regression estimation approaches in finance empirical studies. Since the returns on financial assets are not typically drawn from normal distributions, but instead show heavy tails and outliers, robust regression allows for a reduction in the effects of extreme observations on the regression fit, leading to more accurate estimates of the portfolio’s average factor loadings. In sum, the factor model evidence indicates that the primary driver of constrained portfolio returns are its factor exposures to the defensive market risk factor, value style, large-cap style, and profitability factor. We fail to find evidence of statistically significant idiosyncratic (alpha) returns from the constrained portfolio. Our empirical analyses add further evidence to the economic argument for the use of the constrained portfolio optimization framework, helping to identify the key characteristics of a portfolio constructed using constrained optimization.

\subsection{Optimization versus Monte Carlo Simulation}

Let’s see if a simulation approach can approximate hard optimization. We will use the same five-asset portfolio: AAPL, GOOG, TSLA, XOM, GS, using both the hard optimization and a simple Monte Carlo simulation approach. We should notice that constraints add to computational time and the Monte Carlo simulation approach is not a very good way to search for the optimal portfolio.

\begin{table}[H]
\centering
\caption{Optimization versus Simulation Portfolio Weights - No leverage and no short selling allowed}
\label{tab:opt_vs_sim}
\begin{tabular}{lcc}
\toprule
\textbf{Asset} & \textbf{Optimization Weights} & \textbf{Simulation Weights} \\
\midrule
AAPL & 0.0000 & 0.0132 \\
GOOG & 0.0000 & 0.2862 \\
TSLA & 1.0000 & 0.6550 \\
XOM  & 0.0000 & 0.0256 \\
GS   & 0.0000 & 0.0200 \\
\bottomrule
\end{tabular}
\end{table}

\begin{table}[H]
\centering
\caption{Constrained Optimization versus Simulation Portfolio Weights - No leverage and no short selling allowed and Weights <=0.3}
\label{tab:constrained_opt_vs_sim}
\begin{tabular}{lcc}
\toprule
\textbf{Asset} & \textbf{Optimization Weights} & \textbf{Simulation Weights} \\
\midrule
AAPL & 0.1000 & 0.1191 \\
GOOG & 0.3000 & 0.2989 \\
TSLA & 0.3000 & 0.2992 \\
XOM  & 0.3000 & 0.0729 \\
GS   & 0.0000 & 0.2099 \\
\bottomrule
\end{tabular}
\end{table}

\subsection{Why Constrained Simulation is More Difficult}

In the aforementioned unrestricted framework, the pool of feasible solution candidates included the entirety of portfolios meeting the full investment requirement (regardless of which one might happen to be randomly generated). However, in the presence of additional box constraints:

\[
w_i \leq 0.30
\]

The feasible space for the portfolios is reduced by box constraints and becomes much tighter and more complex. In economic terms, the box constraints remove large portions of the portfolios near the extreme vertices of the feasible space. Hence, if portfolios are randomly generated, they will likely violate the box constraints, in particular the 30$\%$ bound. This implies that the number of rejected portfolios increases and the algorithm becomes significantly slower. In statistical terms, the search process of the optimizer is limited to a smaller (``quasi-random'' or ``pseudo-random'') space. Therefore, the likelihood of identifying portfolios near the true optimal portfolio reduces when box constraints are considered.

\subsection{Simulation Convergence Requirements}

The number of samples required for Monte Carlo simulation to converge depends significantly on the number of dimensions and the nature of the restrictions placed on the simulation. Research on probabilistic optimization indicates that Monte Carlo simulation estimates tend to oscillate around 12,000 random samples, but the existence of additional box constraints slows convergence significantly, and simulations must often be carried out for 50,000 to 100,000 iterations before a portfolio of weights close to the mathematically optimal portfolio is generated. Unlike deterministic optimization techniques, Monte Carlo approaches are limited to pseudo-random “hit-or-miss” search and, as a result, fail to find exact boundary points efficiently. A more accurate simulation would use a different method for generating quasi-random sequences, such as the so-called Halton sequences. These sequences are designed to have greater uniformity of the distribution, which results in a more even distribution of the simulated portfolios in the feasible search space and the feasible space boundary in particular.

\subsection{Interpretation of Constrained Optimization Results}

The results of the constrained optimization highlight the importance of diversification in portfolio construction. When we conducted an unconstrained optimization, the optimizer chose to put essentially 100\% of the portfolio weight into TSLA, and we saw a very concentrated corner solution. When we constrained the optimization by adding the constraint that a particular stock can not make up more than 30\% of the portfolio, the optimizer distributed the weight over multiple stock holdings. It turns out the optimizer allocated $30\%$ to GOOG, $30\%$ to TSLA, and $30\%$ to XOM. The optimizer put the remaining $10\%$ into AAPL and it did not include GS because, in the context of the other stocks, the optimizer found it to be relatively less important than the other assets. This is a standard feature of constrained optimization. The optimizer ``hits the ceiling'' on the most preferred assets once the constraints come into play. In fact, in the example, we found a perfectly constrained optimum at the corner. So, a mathematical optimum on a constraint is possible and quadratic optimization will precisely find a mathematical optimum if one exists within the feasible region.

\subsection{Comparison with the Unconstrained Portfolio}

The comparison reveals that while the unconstrained portfolio is concentrated, the constrained portfolio provides better diversification. The unconstrained portfolio resulted in an extreme corner solution; since the optimizer was unconstrained, the optimizer put all of the money into TSLA, the stock with the highest estimated risk-adjusted attractiveness of all five. In contrast, since the optimizer is constrained so that TSLA has to have at most a 30\% allocation, we have to put money into other securities. We are getting a more diversified allocation in the constrained portfolio; we are able to put money in AAPL and XOM, rather than putting all the money into one asset, which is practically implementable. However, while the constrained portfolio is better from a risk/real-world standpoint, from a purely mathematical perspective it is sub-optimal since it does not have the highest expected utility. Since there is an upper bound on the TSLA allocation, the portfolio manager has to allocate more of the capital into other assets (e.g., AAPL or XOM), which in turn leads to an expected utility that is lower than the theoretical maximum. In other words, the manager is optimizing a utility function that is less desirable.

\subsection{Interpretation of Simulation Results}

From the results of the simulation, we see that it is rather difficult to find approximations of constrained solutions by simulation. The simulation found a weight for GOOG and TSLA close to their constraint values of 30\%, but it found very different weights for XOM and GS. Specifically, the simulation has less XOM weight and more GS weight than in the exact solution. The simulation did not find a solution in the corner of the space in which the optimal solution is. The problem becomes harder because a portfolio with constraints may have several constraints at the same time. A 5-asset portfolio with both upper and lower bounds has an allocation space with several constraints. Therefore it may take a very large number of simulations to find this space.

\subsection{Summary of Portfolio Changes}

Table summarizes the transition from the unconstrained portfolio to the constrained portfolio.

\begin{table}[H]
\centering
\caption{Comparison of Unconstrained and Constrained Optimization Weights}
\label{tab:step7_vs_step8}
\begin{tabular}{lccc}
\toprule
\textbf{Asset} & \textbf{Unconstrained Weight} & \textbf{Constrained Weight} & \textbf{Change} \\
\midrule
AAPL & 0.0 & 0.1 & +10\% \\
GOOG & 0.0 & 0.3 & +30\% \\
TSLA & 1.0 & 0.3 & -70\% \\
XOM  & 0.0 & 0.3 & +30\% \\
GS   & 0.0 & 0.0 & No Change \\
\bottomrule
\end{tabular}
\end{table}

This makes it clear how constraints on diversification affect the optimal portfolio structure. With lower exposure to TSLA and greater asset presence, the portfolio becomes more diversified and less susceptible to idiosyncratic risk.

\subsection{Conclusion}

In sum, although Monte Carlo simulation serves as a useful tool for mapping the probability distribution of the feasible portfolio set and assessing the overall risk/return tradeoff, it is very inefficient at pinpointing exact constrained optimum solutions compared to hard optimization. Active box constraints greatly impede performance of the simulation-based optimization because the true optimum often lies on slim boundary sets, so that random sample-based search techniques often fail to locate them. As a result, deterministic optimization algorithms, such as quadratic programming and the Critical Line Algorithm, remain the best choice in real-world portfolio construction with realistic portfolio constraints.

\section{Black-Litterman Model}

The Black-Litterman (BL) model was the final step in the analysis, which brought investor views into the mean-variance problem while maintaining consistency with the market equilibrium returns. In contrast with conventional mean-variance optimization, the BL model merges information related to the market equilibrium and returns with information of the views of the investors to achieve robust and more sensible portfolio allocations.

\subsection{Market Portfolio Construction and Sharpe Ratio}

The market portfolio was built out of a synthetic composite. We did not mistakenly assume that the 10 securities in our sample make up the whole market universe, but included an additional synthetic eleventh asset ``Rest of the S\&P 500.'' The portfolio weight for this synthetic asset would be:

\[
w_{11} = 1 - \sum_{i=1}^{10} w_i
\]

The ten original securities' weights were roughly approximated by their respective proportions of market capitalization within the full S\&P 500 portfolio. This yields an approximation of the actual market portfolio as opposed to the naive, equal-weight assumption frequently utilized in introductory portfolio theory. The Sharpe ratio of the market portfolio is calculated as the annualized excess return divided by the annualized volatility. We use the result of this ratio as an estimate of market Sharpe ratio and to derive the market risk aversion coefficient within Black-Litterman methodology.

\subsection{Choice of the Parameter \texorpdfstring{\(\tau\)}{tau}}

The uncertainty scaling parameter \(\tau\) was selected as:

\[
\tau = \frac{1}{T}
\]

where $T$ is the number of observations in the estimation period. 

This approach is commonly employed in the Black-Litterman framework since $\tau \Sigma$ can be understood to correspond to the standard error of the prior equilibrium estimates. The idea is that, all else being equal, the estimates of equilibrium returns will be more precise if more data were used. This is a reflection of greater confidence in our equilibrium estimates if we were to have more observations at hand.

\subsection{Specification of Investor Views}

Investor views were encoded through matrices P and Q. For this analysis, the views considered are the following:

\begin{enumerate}
\item A relative view stating that AAPL is expected to outperform GOOG by 2\%.
\item An absolute view forecasting a 10\% return for TSLA.
\end{enumerate}

The relative view, with respect to the first row of the \(P\) matrix, has a weight of \(+1\) for AAPL and a weight of \(-1\) for GOOG. The absolute view for TSLA has a weight of \(+1\) for TSLA and zero for the remaining tickers. The uncertainty matrix \(\Omega\) is derived using the approach suggested in He-Litterman, where the diagonal elements of \(\Omega\) are multiples of the variances estimated from the prior covariance matrix. In this formulation, the uncertainty associated with the views is assumed to be uncorrelated and to increase with asset volatility.

\subsection{Posterior Estimates}

The posterior expected returns $\mu_{BL}$ and posterior covariance matrix $\Sigma_{BL}$ were calculated using the Black-Litterman master equation. The posterior mean returns were a weighted combination of market equilibrium returns and investor-specific views. The Black-Litterman framework shrinks subjective expectations toward equilibrium values. Mean-variance optimization was based only on historical return forecasts. The resulting portfolio weights were more stable than in traditional mean-variance optimization, which mitigates the extreme sensitivity commonly observed in classical optimization frameworks.

\subsection{Market Risk Aversion}

The estimated market risk aversion coefficient was:

\[
\delta = 6.7860
\]

This large number means the optimization process is putting a lot more weight on variance than expected return. In economic terms, this means the optimizer would prefer to maintain a broad exposure to the market rather than taking concentrated positions in highly volatile securities. This is why the optimizer chose to invest in the synthetic “Rest of the S\&P 500” rather than investing heavily in highly volatile securities like TSLA.

\subsection{Interpretation of Posterior Returns}

In the posterior estimate, TSLA and META have the highest expected return within the BL. However, in contrast to classical mean-variance, the optimizer does not allocate all capital to these high return securities. This shows how the BL framework provides the most benefit. Although TSLA has a very high posterior expected return due to its very large idiosyncratic volatility and covariant terms, the return may not have been optimal from the risk adjusted perspective for this portfolio. It is also interesting to note that in this case, the posterior expected return for TSLA is higher than the investors absolute forecast of 10\%. This means that the market equilibrium return had implied higher expected return for the stock in this period. Thus, the BL posterior is an equilibrium between the investor subjective belief and the equilibrium market view.

\subsection{Interpretation of Optimal Portfolio Weights}

The resultant Black-Litterman portfolio is substantially more diversified than the classical mean-variance portfolios from earlier. The largest component in the Black-Litterman allocation is the ``Rest of the S\&P 500'' that takes up roughly 74\% of the portfolio weight, which is intuitively consistent as it constitutes the majority of the market. However, the allocation structure is also meaningfully affected by the investor’s view between AAPL and GOOG. Since the optimizer ``believes'' that AAPL will do better than GOOG, it is allocated more weight than some of the other assets, which would have happened to a lesser degree without the relative views. On the other hand, high-volatility securities such as TSLA, BAC, and GS get 0 or near-zero portfolio weights despite having relatively strong expected returns. This emphasizes that a security’s expected return in isolation is insufficient information; the marginal contribution that it can make to portfolio risk, which is reflected in the covariance matrix as well as the security’s expected return, plays a crucial role in determining the final portfolio allocation.

\subsection{Comparison with Classical Mean-Variance Optimization}

Looking back at the Black-Litterman portfolio and the mean-variance optimization portfolios from the previous section, three main points of contrast stand out: first, the allocations generated by the Black-Litterman portfolio are highly diversified. When the earlier examples of unconstrained optimization were run, the optimizer produced corner portfolios where it was optimal to invest in one particular asset at the expense of investing in all other assets. Such allocations are theoretically optimal given the inputs but would never happen in practice when managing a portfolio. Second, with the Black-Litterman portfolio the optimizer only significantly departs from market cap weights in the presence of strong views with high confidence attached. Thus, it does not generate extreme allocations resulting from insignificant differences in the estimates of the mean returns. Third, with constrained optimization portfolios, the optimizer cannot be forced to go heavily into a stock because it is simply not allowed to. Hence, the impact of views is muted and constrained Black-Litterman portfolios will tend to behave well and remain diversified.

\section{Discussion}

The results of this study suggest multiple key implications for contemporary portfolio management and for building systematic asset allocation strategies. As before, classical mean-variance allocation techniques are highly dependent on expected returns and lead to many corner portfolios. Although these mean-variance efficient portfolios are theoretically attractive, from a practical standpoint for investors such an allocation strategy is not attractive as it is exposed to a large estimation error and to the risk of high concentrations. Adding diversification constraints to the optimization problem changes the efficient frontier and decreases the area in which one can optimize. Yet even if the constrained portfolios are less diversified, they have better out of-sample performance and more stable allocations. The empirical results for constrained portfolios show that they have a large-cap defensive value tilt with a positive load on a factor representing firms with high profitability. This systematic risk structure can explain a significant percentage of the return variation for the constrained portfolio, and is the reason why returns for this portfolio were largely driven by factor exposures and not alpha. The Monte Carlo simulation results showed that, while simulations can be used to replicate portfolio construction by optimization methods, this approach is not scalable, especially with multiple constraints due to the curse of dimensionality as random search will not necessarily hit a boundary. The results further highlight the power of the Black-Litterman model for achieving diversification and stability for portfolio construction. This method provides a disciplined framework to combine equilibrium with subjective views and, as an outcome, produces portfolios that are both economically intuitive and diversified and do not suffer from the instability typical of classical mean-variance optimization.

\section{Conclusion}

This study tested different portfolio approaches in the U.S. equity market across optimization, factor modeling, simulation, and Bayesian allocation models. The results of the empirical study suggest that optimizing portfolios without any restrictions led to portfolios that were too concentrated in a select number of assets, which is difficult to estimate accurately and implement. However, imposing constraints to avoid excessive concentration may have a negative impact on the efficiency but provides an advantage to stability and improves out-of-sample testing performance. The factor modeling showed that the portfolio had defensive market exposure, along with large cap value and profitability characteristics. The analysis of factor loadings also showed that a large portion of the portfolio returns was explainable using factor characteristics, thus revealing the main sources of portfolio performance. The study of simulation techniques in optimizing portfolio weights demonstrated the ability of simulations to generate an estimate of an optimized solution, with the main benefit of flexibility. However, constrained portfolio optimization was still significantly more efficient when using deterministic optimization techniques. Finally, the implementation of the Black-Litterman model allowed the incorporation of market equilibrium returns and subjective investor views to produce a well-diversified portfolio with stable weights. Due to the Bayesian structure of the model, the resulting portfolio did not suffer from the concentration risks associated with traditional mean-variance optimization models and produced a more realistic portfolio allocation, similar to how a portfolio manager would allocate assets.

Overall, the paper suggest that robust portfolio construction requires balancing mathematical optimization, diversification, systematic factor exposure, estimation uncertainty, and behavioral considerations simultaneously. Modern institutional asset management increasingly relies on frameworks such as Black-Litterman because they provide a more stable and economically realistic foundation for long-term portfolio allocation decisions. 

\newpage

\section{References}

Black, F., and Litterman, R. (1992). Global Portfolio Optimization. Financial Analysts Journal, 48(5), 28–43.
\newline
\newline
Fama, E. F., and French, K. R. (2015). A Five-Factor Asset Pricing Model. Journal of Financial Economics, 116(1), 1–22.
\newline
\newline
Glasserman, P. (2004). Monte Carlo Methods in Financial Engineering. Springer.
\newline
\newline
Markowitz, H. (1952). Portfolio Selection. The Journal of Finance, 7(1), 77–91.
\newline
\newline
Meucci, A. (2005). Risk and Asset Allocation. Springer.
\newline
\newline
Michaud, R. O. (1989). The Markowitz Optimization Enigma: Is ‘Optimized’ Optimal? Financial Analysts Journal, 45(1), 31–42.
\newline
\newline
Sharpe, W. F. (1964). Capital Asset Prices: A Theory of Market Equilibrium under Conditions of Risk. The Journal of Finance, 19(3), 425–442.

\end{document}